\documentclass[aps,preprint,showkeys,nofootinbib,superscriptaddress]{revtex4-1}
\pdfoutput=1
\usepackage{amsmath,slashed,amssymb,epsfig,amsthm,bm,graphicx}
\usepackage{graphicx,epsf}
\usepackage{hyperref}
\usepackage[]{latexsym}
\usepackage[top=2.0cm, bottom=2.0cm, left=2.0cm, right=2.0cm]{geometry}

\newcommand{\be}{\begin{eqnarray}}
\newcommand{\ee}{\end{eqnarray}}
\newcommand{\bea}{\begin{eqnarray}}
\newcommand{\eea}{\end{eqnarray}}

\def\comment#1{}

\newcommand{\lp}{\ell_{\rm p}}
\newcommand{\mpl}{m_{\rm p}}
\newcommand{\gn}{G_{\rm N}}

\usepackage{color}

\definecolor{darkred}{rgb}{.8,0,0}

\definecolor{darkblue}{rgb}{0,0,.7}

\definecolor{darkgreen}{rgb}{0,.7,0}

\newcommand{\PoliMilano}{Dipartimento di Matematica, Politecnico di Milano, Piazza Leonardo da Vinci 32, 20133 Milano, Italy}
\newcommand{\Leiden}{Institute-Lorentz for Theoretical Physics, Leiden University, P.O.~Box 9506, Leiden, The Netherlands}
\newcommand{\INFNSalerno}{Dipartimento di Fisica ``E.R. Caianiello",  \\ Universit\`a di Salerno, I-84084 Fisciano (Sa), Italy \& \\
INFN - Gruppo Collegato di Salerno, Italy}
\newcommand{\UCharles}{IPNP - Faculty of Mathematics and Physics, \\ Charles University, V Hole\v{s}ovi\v{c}k\'ach 2, 18000 Prague 8, Czech Republic}
\newcommand{\ELI}{Institute of Physics of the ASCR, ELI Beamlines Project, Na Slovance 2, 18221 Prague, Czech Republic}

\begin{document}
%
\title{Generalized Uncertainty Principle in three-dimensional gravity \\
and the BTZ black hole}
%
\author{Alfredo Iorio}
\email{iorio@ipnp.troja.mff.cuni.cz}
\affiliation{\UCharles}
\author{Gaetano~Lambiase}
\email{lambiase@sa.infn.it}
\affiliation{\INFNSalerno}
\author{Pablo Pais}%
\email{pais@ipnp.troja.mff.cuni.cz}
\affiliation{\UCharles}
\affiliation{\ELI}
\author{Fabio~Scardigli}
\email{fabio@phys.ntu.edu.tw}
\affiliation{\PoliMilano}
\affiliation{\Leiden}


\begin{abstract}
We investigate the structure of the gravity-induced Generalized Uncertainty Principle in three dimensions. The subtleties of lower-dimensional gravity, and its important differences concerning four and higher dimensions, are duly taken into account, by considering different possible candidates for the gravitational radius, $R_g$, that is the minimal length/maximal resolution of the quantum mechanical localization process. We find that the event horizon of the $M \neq 0$ Ba\~{n}ados-Teitelboim-Zanelli micro black hole furnishes the most consistent $R_g$. This allows us to obtain a suitable formula for the Generalized Uncertainty Principle in three dimensions, and also to estimate the corrections induced by the latter on the Hawking temperature and Bekenstein entropy. We also point to the extremal $M=0$ case, and its natural unit of length introduced by the cosmological constant, $\ell = 1 / \sqrt{-\Lambda}$, as a possible alternative to $R_g$, and present a condensed matter analog realization of this scenario.
\end{abstract}
\maketitle
%
%
%
%
\section{Introduction}

The research on the possible modifications of the Heisenberg uncertainty principle (HUP) \cite{Heisenberg1927,Heisenberg1930,Heisenberg1949} has by now a long and established history \cite{Bronstein1936,Bronstein2012,Snyder1947,Yang1947,Mead1964,Karolyhazy1966}. Since the Fourties, many such studies have converged on the idea that some form of generalization of the HUP, usually indicated as Generalized Uncertainty Principle (GUP), must emerge when the effects of gravitation are taken into account. In the last three decades, these generalizations, all resorting to some deformations of the quantization rules, have been proposed in string theory, noncommutative geometry, deformed special relativity, loop quantum gravity, and black hole physics \cite{Amati:1987wq,Gross:1987kza,Amati:1988tn,Konishi:1989wk,MM,Kempf,Dop,FS,Adler2,CLS,SC,Magueijo2002,Arraut2008,Casadio:2013aua,Carr2015}.

As we shall recall below, such gravity-induced GUPs can be extended to \textit{higher dimensions}, $d > 4$, anytime a ``gravitational radius'' (e.g., an event horizon) can be defined. These generalizations have been obtained, for example, in Refs.~\cite{CDM03,Glimpses}. To our knowledge, though, what is still missing is a gravity-induced GUP for \textit{lower dimensions}, $d=3$ and $d=2$. The reasons for this lie in the radically different behavior of key geometric tensors, in lower as compared to higher dimensions. For instance, the Weyl tensor is identically zero in three dimensions, therefore gravitation does not propagate, while the Ricci scalar in two dimensions is just the density of a topological number, the Euler characteristic, hence can carry no dynamics. Such things, that happen when we depart from $d=4$ lowering the dimensions, do not happen when we augment them.

In these days of holography \cite{Bousso2002}, of which the $AdS_3 / CFT_2$ correspondence is a prominent example \cite{Maldacena1999}, lower-dimensional physics is increasingly essential for the theoretical investigation. Also important these days are the analog realizations of high energy theoretical constructions. Examples are the $(2+1)-$dimensional black holes in graphene \cite{IORIO20111334,Iorio_2013,Iorio:2011yz,LongWeylPRDIorioLambiase,IorioReview}, on the one hand, and the GUP stemming from the fundamental length of Dirac materials, on the other hand \cite{Iorio:2017vtw,Iorio:2019cvd} (see also \cite{Jizba:2009qf}). For at least these reasons, it seems an opportune time to fill the gap and build a consistent gravity-induced GUP in lower dimensions.

Our focus will be on three dimensions, where Einstein gravity still makes some sense, and other generalizations of the latter can be naturally included. Furthermore, Einstein gravity with a comological constant, in three dimensions, admit a Ba\~{n}ados-Teitelboim-Zanelli (BTZ) black hole solution \cite{BTZ1992}. On the other hand, the two dimensions are even more unique, as Einstein gravity makes no sense at all, and one has to invent an appropriate theory of gravity from scratch. We shall only briefly comment on this, leaving to a later work a more in-depth analysis.

In what follows we first review, in Section \ref{UPandGrav}, how to achieve a GUP that takes into account the effects of gravitation. In Section \ref{LowDimGUP} we discuss the subtleties involved with the choice of a proper gravitational radius in lower dimensions, especially in $d=2$, and then move to $d=3$ in Sections \ref{newtoniand3} and \ref{BTZGracRadius}, where we focus on the Newtonian gravity, and on the BTZ black-hole, respectively. The latter provides a natural and consistent gravitational radius, hence allows to obtain a GUP. In Section \ref{zeromassbtz} we present a physical realization, in an analog condensed matter system, of the peculiar zero mass BTZ black hole, which will give yet another view on the minimal length. In Section \ref{BTZHawking} we show how the Hawking temperature and Bekenstein entropy of the BTZ black hole are modified when the GUP is taken into account. In the last Section we draw our conclusions, and point to some of the possible future investigations.

\section{Uncertainty Principle in the presence of gravity} \label{UPandGrav}

Let us now briefly review how to achieve a GUP that takes into account the effects of gravitation. One way to do so is to reconsider the argument of the
``Heisenberg microscope'' \cite{Heisenberg1927,Heisenberg1930,Heisenberg1949}: The size
$\delta x$ of the smallest detail of an object, theoretically detectable under such microscope with a beam of photons of energy $E$ (assuming the dispersion relation $E = c p$), is roughly given by
\be
\delta x \simeq \frac{\hbar c}{2\, E} \; ,
\label{HS}
\ee
so that increasingly large energies are required to explore decreasingly small details.

In its original formulation, Heisenberg's \textit{Gedankenexperiment} ignores gravity. However later \textit{Gedankenexperiments} do take it into account, in particular those involving the formation of gravitational instabilities in high energy scattering of strings~\cite{Amati:1987wq,Gross:1987kza,Amati:1988tn,Konishi:1989wk}, or the formation of micro black holes, with an event horizon (gravitational radius), $R_g=R_g(E)$, depending on the centre-of-mass scattering energy $E$, see Ref.~\cite{FS}. Such scenarios suggest that (\ref{HS}) should be modified to
\be
\delta x \simeq \frac{\hbar c}{2\, E} + \beta\, R_g(E) \;,
\label{GUP}
\ee
where $\beta$ is a dimensionless parameter, and $R_g$ is the gravitational radius associated with $E$. The deformation parameter $\beta$, in principle, is not fixed by the theory, although it is generally assumed to be of order one. This happens, in particular, in some models of string theory (see again, e.g., Refs.~\cite{Amati:1987wq,Gross:1987kza,Amati:1988tn,Konishi:1989wk}), and has been confirmed in Ref.~\cite{SLV} where an explicit calculation of $\beta$ has been performed. A lively debate is however present in literature on the "size" of $\beta$ (see for example Refs.~\cite{ScardDefPam,ScardGravTest,Luciano2019,Pikovski2011,PlenioKumar2019,Villalpando2019,Bushev2019,Bosso2018,Feng2016}).

In $d=4$ dimensions\footnote{The Planck length is defined as $\lp=\sqrt{\gn\,\hbar/c^3}\simeq 10^{-33}\,$cm, with $\gn$ the Newton constant. The Planck energy is ${\cal E}_p = \, \hbar\, c/(2 \lp)$, and the Planck mass is $\mpl= {\cal E}_p /c^2$. The Boltzmann constant $k_{\rm B}$ will be shown explicitly, unless otherwise stated.} $R_g\ = 2\, \lp^2\, E / (\hbar c)$, hence (\ref{GUP}) becomes
\be
\delta x \simeq \frac{\hbar c}{2\, E} + 2\beta\, \lp^2\frac{E}{\hbar c} \;.
\label{He}
\ee
This kind of modification was also proposed in Ref.~\cite{Adler2}.

\par
Relation~(\ref{He}) can be recast in the form of a GUP ($\delta x \to \Delta x$, $E \to c \Delta p$ and $\lp = \hbar / (2\mpl c)$)
\be
\Delta x\, \Delta p \geq \frac{\hbar}{2} \left[1+\beta\left(\frac{\Delta p}{\mpl c}\right)^2\right] \; .
\label{gup}
\ee
For mirror-symmetric states (with $\langle \hat{p} \rangle = 0$), since $\Delta x\, \Delta p \geq (1/2)\left|\langle [\hat{x},\hat{p}] \rangle\right|$,
the inequality (\ref{gup}) implies the commutator
\be
\left[\hat{x},\hat{p}\right] = i \hbar \left[1+\beta\left(\frac{\hat{p}}{\mpl c} \right)^2 \right] \; .
\label{gupcomm}
\ee
Vice-versa, the commutator (\ref{gupcomm}) implies the inequality~(\ref{gup}) for any state.
The GUP is widely studied in the context of quantum mechanics~\cite{Vagenas,Ong2018,Nozari}, quantum field theory~\cite{Husain:2012im,Lambiase:2017adh,Lake2018}, thermal effects in QFT~\cite{FS9506,AdSTemp,Gine2018,Scardigli:2018jlm,Buoninfante2019,Giusti2019}, and for lattice formulation of the quantization rules~\cite{Jizba:2009qf}.

A couple of comments are now in order. The gravitational radius appearing in formula (\ref{GUP}) has been initially introduced for spherical symmetric situations, in particular the Schwarzschild case for $d\geq 4$. While, for the sake of simplicity, the use of spherical symmetry can be justified here,  relation (\ref{GUP}) certainly might enjoy future improvements to the non spherical case. A similar fate was that of the original Bekenstein bound, with the emergence of a characteristic radius that, over the years, enjoyed modifications to the spherical symmetric formula (see, e.g., Bousso review \cite{Bousso2002}).


Another comment is that the GUP stemming from strings or micro black holes \textit{Gedankenexperiments} is substantially different from the approach of noncommutative geometry (see, e.g., \cite{Dop} and also \cite{ioriotheta,LambiaseKanazawa2019}). While there a general commutator $[x_\mu,x_\nu]=i\hbar\theta_{\mu\nu}(x)$ is postulated on the grounds of non commutative geometry insights, here we introduce a commutator dictated essentially from high energy scatterings, re-examined in specific \textit{Gedankenexperiments}. Further connections and comparison with the approach of Ref.\cite{Dop} will be discussed in future works.

As mentioned, the formula (\ref{GUP}), and the related GUP, can be easily generalized to $d > 4$, anytime $R_g$ can be defined~\cite{CDM03,Glimpses}. Let us show now how to proceed when $d = 2,3$.


%
%
%
%
%
%
%
%
%

\section{Lower dimensional GUP} \label{LowDimGUP}

The main message of the previous Section is that the existence of a gravitational radius affects the localization, as expressed in formula (\ref{GUP}). We shall assume that a version of that formula is also valid in lower dimensions, as long as a gravitational radius can be identified. In what follows we shall discuss several options.

A fundamental observation is that, for $d = 2, 3$, Einstein gravity, and the corresponding Newtonian limit, decouple. Hence, we are lead to three possibilities: \\ (A) To develop a coherent Newtonian gravity in $d=2$ or in $d=3$ dimensions. These, in general, cannot be derived as limits of Einstein gravity;\\
(B) To rely on Einstein gravity (perhaps, including a cosmological constant) at least for $d = 3$;\\
(C) To go beyond Einstein gravity, either ($d=3$) by adding to the Einstein-Hilbert (EH) term other admissible terms, such as the Chern-Simons gravitational term, see, e.g., \cite{DESER1982372} and \cite{GURALNIK2003222}, or ($d=2$) by proposing entirely new dynamical models, often based on scalar fields (dilatons), see, e.g., the review \cite{GKV}.

In the next Section we shall focus on $d=3$ by elaborating on the cases (A) and (B), since in these cases there is a clear $d=4$ correspondence, while case (C) deserves a separate later study. But before going there, let us only briefly comment on $d=2$.

As well known, the EH action in two dimensions amounts to a topological number
\be
\int_{\cal M} \sqrt{g}\, R\, d^2x \ = \ 2 \pi \chi \; ,
\ee
where $\chi$, the Euler characteristic, depends only on the topology of the spacetime manifold $\cal M$. As a consequence, the Einstein tensor identically vanishes. Henceforth, one needs to invent from scratch a suitable theory, whose dynamics plays the role of Einstein field equations. This opens the doors to a variety of candidates for two-dimensional gravity, as one can see by combing through the references of \cite{GKV,GM}. Two-dimensional black holes, with their temperatures, entropies and the whole thermodynamics, can be defined for some of these theories, see \cite{BHT, evanescent_black_holes, Witten1991}, and also the recent \cite{MPW}. However, in this lineal world it is not clear whether it makes sense to talk about a consistent $R_g$. The meaning of $R_g$ itself is, of course, model dependent, just like the specific gravity one uses for its definition. In other words, the  $d=2$ world needs a separate study, for each black hole stemming from a specific gravity model. It is surely worth it, but we shall not perform that here. We want, instead, merely point to the complexity of this case, and move to the more tractable case of $d=3$, first considering the Newtonian gravity and then the Einstein gravity.

\section{$d=3$ Newtonian gravity and inconsistent $R_g$}\label{newtoniand3}

As said above, in $d=3$ (and in $d=2$) Einstein gravity does not have a straightforward Newtonian limit, opening the doors to many different speculations \cite{Romero-Dahia}. In this case, the reason is that in three dimensions the Weyl tensor, responsible of the non-trivial solution of the Einstein field equations, outside a matter region ($R_{\mu\nu}=0$), identically vanishes.

To develop Newtonian gravity we require the validity of Gauss theorem, also in $d=3$. Then the Newtonian gravitational field, $\vec{g}$, of a point mass $M$ should be
\be
\vec{g}=-\frac{GM}{r^{2}} \, \vec{r}
\ee
so that the flux through the circle $S = 2 \pi r$ is
\be
\Phi_S(\vec{g}) = \frac{GM}{r} \cdot 2\pi r = 2\pi GM \;.
\ee
Notice that here $G$ cannot be the usual Newton constant of $d=4$, $\gn$. However, if we demand that the field $\vec{g}$ has the dimensions of an acceleration, $[g]=L/T^{2}$, then the product $GM$ should have the dimension of a speed squared, $[GM]=L^{2}/T^{2}$. Comparing the latter with the $d=4$ result, $[\gn M]=L^{3}/T^{2}$, we see that, if we want to keep as fundamental the dimension of a mass, $M$, then
\be
[G] = \frac{[\gn]}{L}\,.
\ee
This way, the fundamental dimensions of length, time and mass, are preserved in $d=3$, just as in $d=4$.

\begin{figure}
\begin{center}
\includegraphics[width=0.4\textwidth,angle=0]{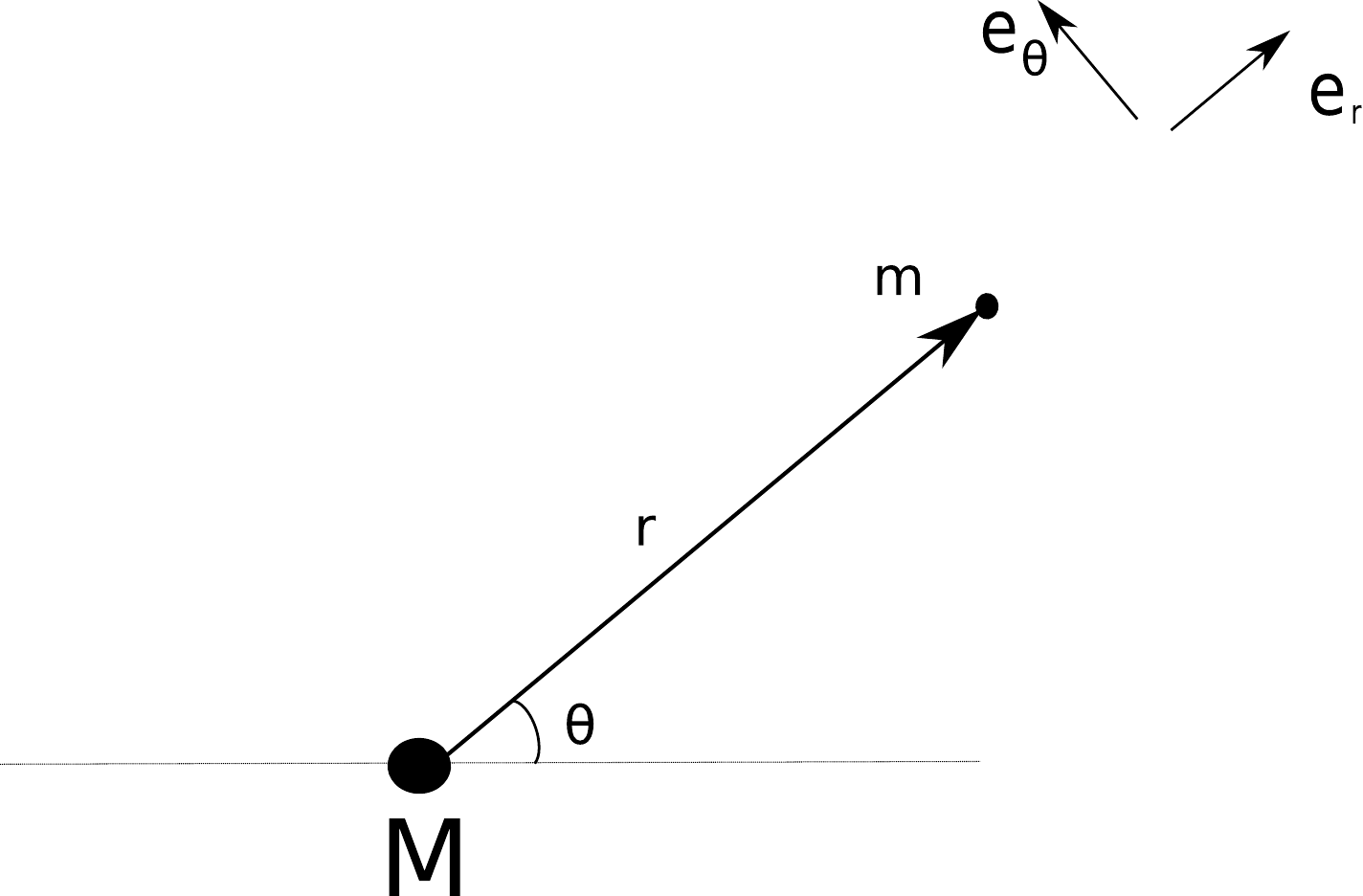}
\end{center}
\caption[3pt]{{\protect\small {In the text we consider the effective potential $V_{eff}$ of a configuration with a very large mass $M$ interacting  with a point-like mass $m$ ($M\gg m$).}}}%
\label{masses_fig}%
\end{figure}

\begin{figure}
\begin{center}
\includegraphics[width=1.0\textwidth,angle=0]{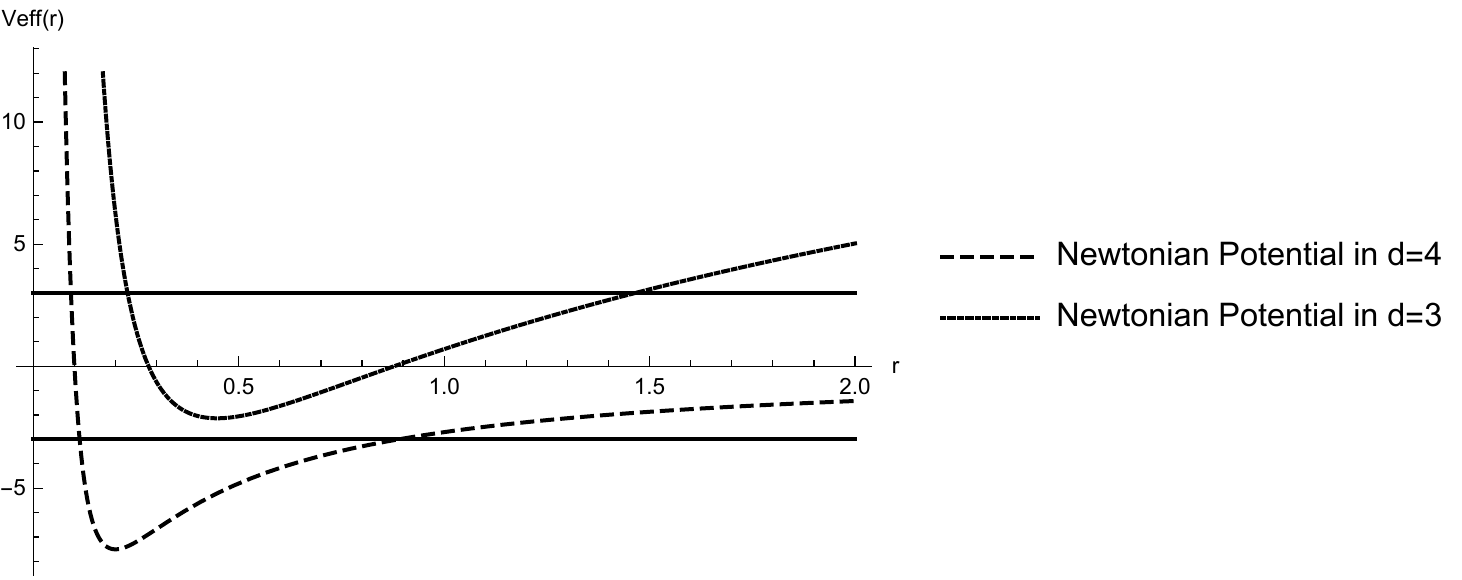}
\end{center}
\caption[3pt]{{\protect\small {The Newtonian gravitational effective potential in $d=3$ (short dashes) and in $d=4$ (long dashes). The horizontal continuous lines, that refer to arbitrary values of the total energy $E$, help visualize that, for any value of the allowed energies $E$, the orbits in $d=3$ are always bounded, i.e. the value of $r$ can never exceed a value fixed by the intersection of $V_{eff}$ and the given horizontal line (here, roughly given by $r \simeq 1.5$ for $E=+3$).}}}%
\label{orbitals_fig}%
\end{figure}

\begin{figure}
\begin{center}
\includegraphics[width=0.7\textwidth,angle=-90]{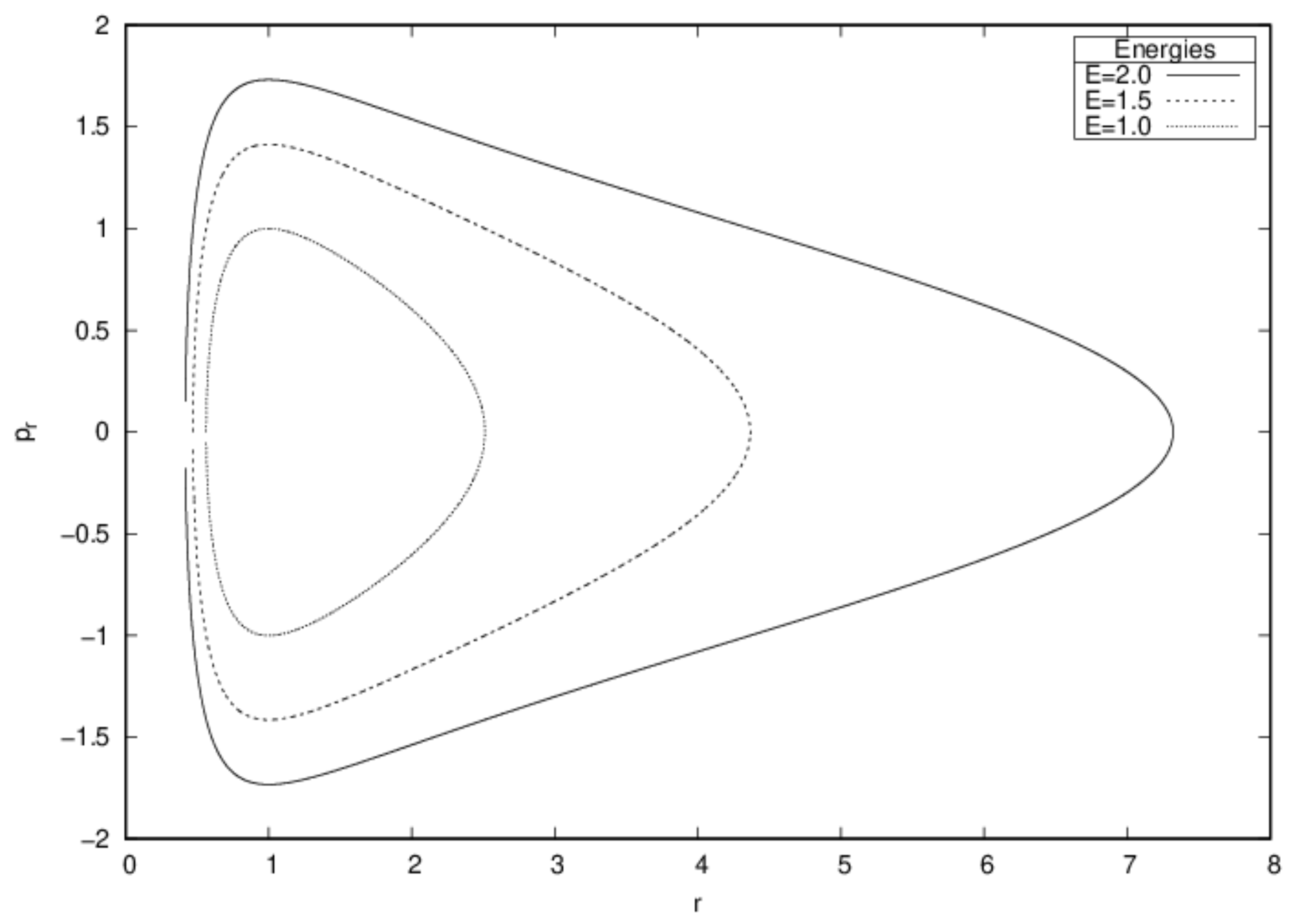}
\end{center}
\caption[3pt]{{\protect\small {The trajectories of a particle of mass $m=1$ and $l=1$ in $d=3$. The plots are in the phase space of the radial coordinate, $(r, p_r)$, for three different values of the energy, $E=1,1.5, 2$. As discussed in the text, this illustrates the inescapable bounded nature of the orbits of massive particles in $d=3$ Newtonian gravity.}}}%
\label{phase_space_fig}%
\end{figure}

The gravitational potential then reads
\be
\label{potential_2+1}
V(r)=GM\ln(r/r_{0}) \;,
\ee
where $r_{0}$ identifies the zero of the potential, $V(r_{0})=0$. Notice the positive sign on the right hand side of (\ref{potential_2+1}), that gives the gravity field the correct direction
\be
\vec{g} = - \vec{\nabla} V = -\frac{GM}{r^{2}} \, \vec{r}\,.
\ee
To identify a possible gravitational radius let us introduce an effective potential, $V_{eff}$, and analyze its behavior. We consider a particle of mass $m$, at radial distance $r$ from a much larger mass $M\gg m$ (see Fig.\ref{masses_fig}), and suppose that the gravitational potential, $V(r)$, generated by $M$ is as in
\eqref{potential_2+1}. Then the gravitational potential energy of the system is $U=mV$, and from the Lagrangian
$$\mathcal{L}=T-U=\frac{1}{2}m\dot{r}^{2}+\frac{1}{2}mr^{2}\dot{\theta}^{2}-GMm\ln(r/r_{0})\;,$$
we obtain the equations of motion \cite{Goldstein,Landau-Lifshitz}
\begin{eqnarray}
mr^{2}\dot{\theta}&=&j=\mbox{constant} \;,\\
m{\ddot{r}}&=&\frac{j^{2}}{mr^{3}}-GM\frac{m}{r} \nonumber \;.
\end{eqnarray}
Integration of the second equation, leads directly to the energy
\be
\label{eqnergy_conservation_2+1}
E=\frac{1}{2}m\dot{r}^{2}+ GMm\ln(r/r_{0})+ \frac{j^2}{2mr^{2}}\;,
\ee
that is always bounded from below, $E \ge GMm\ln(r/r_{0}) + j^2/(2mr^{2})$, otherwise $\dot{r}$ would be imaginary. This allows to define the wanted effective potential as $U_{eff} =  m V_{eff} \equiv m \left[ GM \ln(r/r_{0}) + j^2/(2 m^2 r^2) \right]$.
In Fig.\ref{orbitals_fig} we see the consequences of this. For any allowed value of the total energy (e.g. $E= \pm 3$ in the figure), the particle's orbit must be bounded (closed), as can be seen also in Fig.\ref{phase_space_fig}.

This should be compared with the $d=4$ case. There, if the total energy is bigger than some value (in general set to zero), the orbit is not bounded. Therefore, the point-like particle can escape to infinity. On the contrary, in $d=3$ the logarithmic behavior of $V_{eff}$ at $r\to+\infty$ makes the orbits bounded, no matter how big the total energy $E$ is.
As well known, in $d=4$, this allows for a clean definition of a gravitational radius: One needs to consider the first unbounded orbit at $E=0$, and define an ``escape velocity'' $v_f$, as the velocity necessary to a point particle to escape from a distance $r$, from $M$, to infinity
\be
v_f^2 = \frac{2\gn M}{r} - \frac{j^2}{m^2 r^2} \;.
\ee
For a radial path (i.e., for $j=0$), and considering the limiting case of $v_f \to c$, we obtain the wanted gravitational radius from $c^2 = 2\gn M/R_g$, that is $R_g = 2 \gn M/c^2$.

The same steps cannot be repeated in the $d=3$ case, simply because there are no unbounded orbits, i.e. all the orbits are closed, and therefore there is no escape velocity. Thus, our suggestion here is simply
\begin{equation}
\label{RgNewLimit}
R_g = {\rm undefined}\,.
\end{equation}
Of course, when light is seen as a bunch of photons, that are relativistic massless particles, Newtonian gravity cannot affect them. In that sense, a black-hole cannot even be defined in a consistent way. On the other hand, if we take the old Newtonian view of light as particles with tiny mass, we could say that the radius of the black hole horizon in $d=3$ Newtonian gravity is infinite. These arguments about light, though, are better faced in a fully relativistic approach. This, and the previous arguments, make us move to the next Section, to keep searching for a consistent $R_g$.

\section{BTZ black hole, consistent $R_g$ and the GUP}  \label{BTZGracRadius}

Given the previous puzzling results, that do not allow to define a consistent gravitational radius in $d=3$ Newtonian gravity, we consider here, instead, Einstein gravity with a cosmological constant: $\int d^3 x \sqrt{g} (R - 2 \Lambda$). Indeed, this is probably the most direct way to proceed, that is, to simply write the $d=4$ action in $d=3$, and {\it define} that to be the $d=3$ theory of gravity.

In what follows, we shall discard the case $\Lambda > 0$, which furnishes a natural (de Sitter) radius, that is the location of the cosmological horizon. Such horizon cannot be identified with the $R_g$ we are looking for, because it has nothing to do with the process of measurement and quantum localization of a particle, that we discussed at length in the first two Sections of this paper. On the other hand, when $\Lambda < 0$, the theory supports the well known BTZ black hole solution, with a proper event horizon that can naturally be associated to the wanted $R_g$ (see Refs.\cite{BTZ1992,BHTZ,Mann1993,Mann1995}).

To write the metric describing the BTZ black hole in ``Schwarzschild coordinates'', we follow here Ref.\cite{Carlip2005}, with some small changes. In particular, we work with $c \neq 1$. Moreover, although Einstein gravity in $d=3$ dimensions does not have a Newtonian limit, we want to keep some contact with Newton theory. Therefore we choose the parameter $M$ to measure a physical mass, and the gravitational constant $G$ to be the same as in $d=3$ Newtonian theory. Hence, as before, $[GM] = L^2/T^2$. With these conventions the BTZ metric reads \cite{BTZ1992,Carlip2005}
\be\label{btzgeneral}
ds_{BTZ}^2 = f(r)^{2} c^2dt^2 - f(r)^{-2}dr^2 - r^2(d\phi + N^\phi cdt)^2
\ee
where
\be
f^2(r) = -\frac{8GM}{c^2} - \Lambda r^2 + \frac{16G^2J^2}{c^4\,r^2}\,, \quad \quad \quad
N^\phi=-\frac{4GJ}{c^2\,r^2}\,,
\label{btzcarlip}
\ee
where $M$ is the mass (the conserved charge associated with the asymptotic invariance under time displacements), and $\Lambda<0$ is the negative cosmological constant, as said earlier. Furthermore, $J$ is the conserved charge associated with rotational invariance, namely the angular momentum. As usual (see, e.g., \cite{Walker}), horizons are located at the positive zeros of the function $f(r)$. In this case they are two, $r_+$ and $r_-$, given by
\be\label{horizonbtz}
r^2_\pm = \frac{4GM \ell^2}{c^2}\left[1 \pm \left(1 - \frac{J^2}{\ell^2 M^2} \right)^{1/2}\right] \,.
\ee
where, from now on, we write $\Lambda \equiv - 1/\ell^2 < 0$

We have a black hole under the conditions
\be
M> 0 , \quad \quad \quad \quad \quad |J| \leq M \ell
\ee
with $r_+$ a genuine event horizon, and $r_-$ a Cauchy horizon (when $J\neq 0$). There also exist solutions with other values of $M$ and $J$, which are not black holes but conical naked singularities, discarded on physical grounds. There is, though, an important exception that is the case $M=-1$ (in units where $8G/c^2=1$) and $J=0$, which corresponds to the Anti-de Sitter space \cite{Miskovic-Zanelli,BHTZ}. The latter solution indicates that the "vacuum state", namely the extremal case $M \to 0$, which implies $J \to 0$ too, is not the bottom of the spectrum, but rather a peculiar ``massless black hole'', whose (empty) spacetime has line element
\be \label{zerobtz}
ds_0^2 = (r/\ell)^{2} c^2 dt^2 - (r/\ell)^{-2}dr^2 - r^2 d\phi^2 \;.
\ee

Therefore, even in the extremal case of a ``massless BTZ black hole'', one can introduce a special value of $r$, that is $r=\ell$, that is a sort of natural unit of length. Of course, this does not make $r=\ell$ an event horizon, as such, but further physical inputs are necessary to use $\ell$ as the minimal length of quantum localization we are seeking. In the next Section we shall present a condensed matter analog realization of this scenario. There, the physics of $\ell$ indeed is clear, and points to a fundamental length. Before that, let us focus on the general case of a gravitational radius associated to nonzero $M$.

For simplicity, we keep spherical symmetry, that is we choose $J=0$, so that a natural $d=3$ gravitational radius can eventually be defined as
\be
R_g \equiv r_+ = \frac{\ell}{c} \, \sqrt{8GM}\,.
\label{rg2}
\ee
We shall soon build on this definition to obtain the GUP formula we are looking for. Before doing so, we present an argument about the BTZ black hole formation mechanism.
In Ref.\cite{CruzZanelli} it is shown that a gravitational collapse, that ignites the black hole formation, is best obtained for a perfect fluid. For point-like masses things are different, because in three dimensions gravity does not propagate, and the point-like mass just creates a conical singularity \cite{Star,tHooft}. In $d=3$ Einstein gravity the formation of a non rotating black hole horizon is impossible, without a negative cosmological constant.

For the perfect fluid, according to the results of \cite{CruzZanelli}, the formula (\ref{rg2}) for $R_g$ should be modified to
\be
\label{RgBTZgamma}
R^{(\gamma)}_{g} = \frac{\ell}{c} \, \sqrt{8 G' (M - \gamma)}
\ee
where $\gamma$ is a constant that depends on the perfect fluid, $G'$ is a constant with the dimensions of the Newton constant in $d=3$ that needs not be the same as the $G$ of the previous discussion (since, as we know, the Newtonian limit does not necessarily apply here).

Having said that, for the sake of both simplicity and generality, we stick here to the formula (\ref{rg2}), and we leave to future analysis the discussion about the physical formation of a $d=3$ black hole. Hence, considered the energy $E$ involved in the scattering process of the localization measurement, and the equivalent mass $M \to E/c^2$ of the ensuing micro BTZ black hole, then we can write
\be \label{chosenRg}
R_g(E) = \frac{\ell}{c^2} \, \sqrt{8GE}  \;,
\ee
and the $d=3$ version of the minimal spatial uncertainty (\ref{GUP}) reads
\be
\delta x \simeq \frac{\hbar c}{2\, E} + \beta\, \frac{\ell}{c^2} \, \sqrt{8GE} \; .
\label{2+1}
\ee
Following standard procedures (see, e.g., Refs.\cite{FS,Adler2,ACSantiago,Susskind}), and assuming the dispersion relation $E=pc$ (in general valid for any high energy particle), a little algebra allows to recast (\ref{2+1}) into a deformation of the uncertainty principle
\begin{equation}
\Delta x\, \Delta p \geq \frac{\hbar}{2} \left[1 +  4\beta\sqrt{\frac{2G\ell^2}{\hbar^{2}}\left(\frac{\Delta p}{c}\right)^{3}}\right]\ .
\label{3gup}
\end{equation}
Note that the second term in the squared brackets is dimensionless, as it must be. Furthermore, it is possible to define a $d=3$ Planck mass as
\be
m_{p} \equiv \sqrt[3]{\frac{\hbar^{2}}{2 \ell^2 G}} \; .
\label{pl}
\ee
With this, equation \eqref{3gup} becomes our GUP in $d=3$, and can be written as
\begin{equation}
\Delta x\, \Delta p \geq \frac{\hbar}{2} \left[1 +  4\beta\left(\frac{\Delta p}{m_{p}c}\right)^{3/2}\right]\ .
\label{3gup_pl}
\end{equation}


Note that, in this case it is not straightforward to write a commutator which implies the inequality (\ref{3gup_pl}). We have been able to do so for equations (\ref{gup}) and
(\ref{gupcomm}) because, for any given operator $\hat{A}$, we could use the equality $(\Delta A)^2 = \langle \hat{A}^2 \rangle - \langle \hat{A} \rangle^2$. Here the different exponent, $(\Delta A)^{3/2}$, does not allow to write a similar expression. Finally, in the limit $\beta \to 0$, we recover the standard HUP, $\Delta x\, \Delta p \geq \hbar/2$.


\section{Condensed matter analog of $M=0$ BTZ and $\ell$ as minimal length} \label{zeromassbtz}

Let us now present the promised condensed matter example of an analog of a zero mass BTZ black hole, where there is a natural physical interpretation of $\ell = 1 / \sqrt{-\Lambda}$ as the minimal length of the system.

The system we refer to is a two (spatial) dimensional Dirac material \cite{WehlingDiracMaterials2014}, a prototypical example being graphene \cite{PacoReview2009}. Indeed, it is by now about a decade that, due to their low energy spectrum, Dirac materials  have emerged as powerful condensed matter analogs of high energy phenomena \cite{IORIO20111334,Iorio_2013,Iorio:2011yz,LongWeylPRDIorioLambiase,IorioReview,Iorio:2019cvd,Iorio:2015iha,Iorio:2017vtw,IorioPais}. In particular, in \cite{LongWeylPRDIorioLambiase} analogs of Dirac quantum fields on a variety of graphene spacetimes with nontrivial \textit{curvature} have been proposed (see also the open debate on spacetimes with nontrivial \textit{torsion} \cite{Mesaros:2009az,IORIO2018265,iorio2019torsion}). Particularly important for us here are two aspects of that research: one is the BTZ of \cite{LongWeylPRDIorioLambiase}, and one is the emergence of a GUP from the lattice constant, the length scale of the material
\cite{Iorio:2017vtw,Iorio:2019cvd,Jizba:2009qf,Jizba:2013aj}.

\begin{figure}[tbp]
\begin{center}
\includegraphics[width=0.7\textwidth]{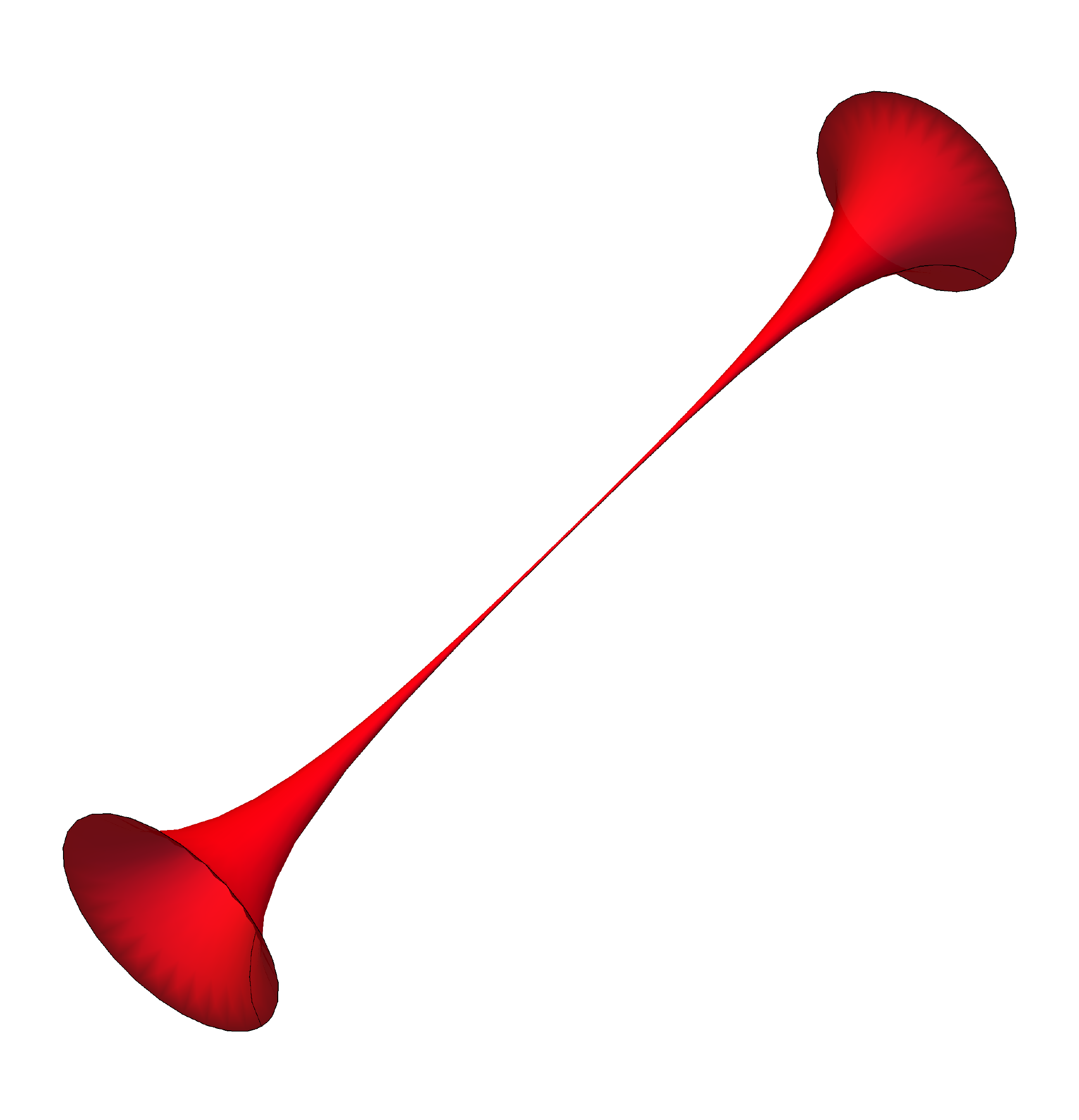}
\end{center}
\caption{\label{fig2} The hyperbolic pseudosphere for $a=1$, $C=1/100$. Clearly for $C/a \to 0$, the surface tends to two Beltrami pseudospheres joined at the minimum value of $\rho$, that is $\rho_{\rm min}= C$. In the plot, the ``Hilbert horizons'' are two, and located at the two maximal circles $\rho_{\rm max} = \sqrt{a^2 + C^2} \simeq 1.00005$.} \label{hyperbolicwormhole}
\end{figure}

In \cite{LongWeylPRDIorioLambiase} it was shown that the metric of the $J=0$ BTZ black hole is conformal to the metric of a spacetime $\Sigma_{\rm HYP} \times \mathbf{R}$, where the spatial part, $\Sigma_{\rm HYP}$, is the hyperbolic pseudosphere \cite{Eisenhart1909}, see Fig.\ref{fig2}, while $\mathbf{R}$ is spanned by time. One important point here is that the hyperbolic pseudosphere belongs to the family of surfaces of constant negative Gaussian curvature
\be
K = - 1/a^2 \,.
\ee
As such, since a real lab is in $\mathbf{R}^3$, such surfaces can only represent portions of the Lobachevsky plane, hence necessarily have boundaries, cusps, self-intersections, or other kinds of singularities, as established by a theorem of Hilbert, see, e.g., \cite{ovchinnikov}. In particular, since the surface in point is a surface of revolution,  with line element
\be
dl^2 = du^2 + C^2 \cosh^2(u/a) d\phi^2 \,,
\ee
with $u$ the logitudinal coordinate, and $\phi \in [0,2 \pi]$, the locus of such singular boundary is a circle. In terms of the radial coordinate
\be \label{radiusHyp}
\rho(u) = C \cosh(u/a) \,,
\ee
such circle is the maximal, $\rho_{\rm max} = \sqrt{a^2 + C^2}$, where $C$ is the minimum, $\rho_{\rm min} = C$, cf. Fig.\ref{fig2}.

As a tribute to Hilbert, and with a little abuse of the word ``horizon'', such locus in \cite{LongWeylPRDIorioLambiase} has been called ``Hilbert horizon'', $\rho_{\rm max} = \rho_{Hh}$. In fact, it is not an horizon in the general relativistic sense. On the other end, it is not even a boundary one is free to move, as for the cylinder, or to remove, as for the sphere (for a general introduction to the latter case, see the classic \cite{Eisenhart1909}, while for a recent application, closer to the present discussion, see \cite{ioriokus2020}).

Knowing this, one could conclude that, in general, the Hilbert horizon and the event horizon could not match, as noticed in \cite{CVETIC20122617}. For a non-extremal hyperbolic pseudosphere, strictly speaking, this is true. Nonetheless, when the role of the $C$ parameter is duly taken into account, the two horizons can be meaningfully made to coincide in the $C / a \to 0$ limit. The mass of the hole goes to zero even faster, hence we have the $M \to 0$ BTZ we announced. In that limit the hyperbolic pseudosphere tends to two Beltrami pseudospheres ``glued'' at the tails, as shown in Fig.~\ref{fig2}.  Let us show this here.

Let us rewrite the line element of the BTZ black hole in (\ref{btzgeneral}), setting to zero the angular momentum in (\ref{btzcarlip}), and easing a little the notation by setting\footnote{This hides important issues about the physical meaning of the ``speed of light'' $c$ here, but has the advantage of focusing entirely on the role of length scale $\ell$. On the importance of $G$ in this context we extensively commented earlier.} $8 G / c^2$ to 1. With this
\begin{eqnarray}
    ds^2_{BTZ} &= & \left(r^2/\ell^2 - M\right) dt^2
    - \frac{dr^2}{\displaystyle{r^2/\ell^2 - M}}-r^2 d\phi^2 \nonumber \\
    &\equiv &\left(r^2/\ell^2 - M \right) ds^2 \;, \label{eq1}
\end{eqnarray}
where, as we know, $\Lambda = - 1/\ell^2 < 0$,
\begin{equation}\label{eq4}
    ds^2\equiv  dt^2-\ell^4\frac{dr^2}{\displaystyle{\left(r^2-r_+^2\right)^2}} - \ell^2\,
    \frac{r^2}{\displaystyle{\left(r^2-r_+^2\right)}} d\phi^2\,,
\end{equation}
and
\be\label{hypevent}
r_+ \equiv \ell \sqrt{M} \;,
\ee
as in (\ref{horizonbtz}), adapted to this case ($J=0$) and to this notation.

Let us define
\begin{equation}\label{eq7}
    du \equiv - \frac{\ell^2}{r^2-r_+^2}\, dr \,, \qquad \rho(r) \equiv \frac{\ell r}{r^2-r_+^2}\,,
 \end{equation}
from which one obtains
\be\label{radiusBTZ}
r (u) = r_+ \coth(r_+ u / \ell^2) \;,
\ee
that gives
\be
\rho(r (u)) \equiv \rho(u) = \ell \cosh(r_+ u / \ell^2) \;.
\ee
Comparing the latter with (\ref{radiusHyp}), we see the hyperbolic pseudosphere, with the $C$ parameter (the smallest radius $\rho$) equal to the ``cosmological'' parameter
\be \label{Cell}
C  \equiv \ell \,,
\ee
and the radius of curvature, $a$, related to the former parameter and to the radius of the event horizon
\be \label{aANDr+}
a \equiv \ell^2 / r_+ \,.
\ee
With this, one sees that the line element in (\ref{eq4}) is that of $\Sigma_{\rm HYP} \times \mathbf{R}$, so that
\be\label{btzhyp}
ds^2_{BTZ} = \left(r^2/C^2 - M \right) ds^2_{\rm HYP} \;,
\ee
with
\be
M = C^2 / a^2 \,.
\ee
The last formula is obtained by using (\ref{Cell}) and (\ref{hypevent}) in (\ref{aANDr+}).

We then need to notice that, in a laboratory realization of the structure in Fig.\ref{fig2}, the narrowest throat of the pseudosphere, corresponding to $\rho_{\rm min} = C$, cannot have radius smaller than the lattice constant of the given Dirac material, otherwise the structure would break. This simple and evident argument makes our point here. That is, the physical meaning of $C$, hence in turn of $\ell$, is the lattice constant, $\ell_L$, the most natural minimal length of the system
\be
\ell = C = \ell_L \,.
\ee
Of course, the last equality is an idealization, and only holds approximately, as such structures in real lab, for stability, require a bigger $\rho_{\rm min}$ (for the case of graphene see \cite{iorioWeylLab2012,Taioli_2016}).

Therefore, the BTZ black hole relevant quantities, after this identification, are given by
\be\label{identifBTZ}
\Lambda \equiv - 1 / \ell_L^2 \quad , \quad M \equiv \ell_L^2 / a^2 \quad , \quad r_+ \equiv \ell_L^2 / a \;.
\ee

Let us now compare the event horizon, $r_+$, to the Hilbert horizon of the hyperbolic pseudosphere spacetime
\be\label{realBTZhorizon}
\rho_{H h} = \sqrt{a^2 + \ell_L^2} = a \, \sqrt{1 + \ell_L^2 / a^2} \;,
\ee
which is given in different coordinates, though. This is easily obtained if we use the corresponding meridian coordinate, $u_{H h} = a \, {\rm arccosh} \left( \sqrt{1 + a^2/\ell_L^2} \right)$, substitute this value into (\ref{radiusBTZ}), and use (\ref{identifBTZ})
\be
r_{H h} \equiv r (u_{H h}) = r_+ \coth\left( {\rm arccosh} \left( \sqrt{1 + a^2/\ell_L^2} \right) \right) \;.
\ee
For $a = 10^n \ell_L$ this formula approximates to
\begin{equation}\label{eqrn}
    r_{H h} = r_+ \times \frac{10^n}{(10^{2n}-1)^{1/2}}\simeq r_+ \times \left( 1 + 5\times 10^{-(2n+1)} \right) \,.
\end{equation}
Clearly, in the limit of small $\ell_L/a$, these two horizons coincide. That is also the limit where $M \to 0$, and, accordingly $r_+ \to 0$, i.e. the zero mass black hole we have announced, or what in \cite{BTZ1992} is called ``the vacuum state''.

The spectrum of the BTZ is continuous from $M=0$ on, for growing values of the mass, $M>0$. As said earlier, this continuous spectrum corresponds to black holes, the extremal case being $M \to 0$. Between $M=-1$ and $M=0$ the spectrum is discrete, and corresponds to conical singularities. The AdS is reached only when $M=-1$, that is the true end of the spectrum. Therefore, one may say that there is still ``something of the black hole'', even in the $M=0$ case. This is in contrast with the higher dimensional case, where at $M = 0$ all features of the black hole are gone. So, in this context we may as well choose to define
\be
R_g \equiv \ell_L \;.
\ee
The logic of this choice is that we learnt of this ``radius'' when dealing with a gravitational object, that is the $M=0$ BTZ black hole. Nonetheless, its meaning is somehow deeper than the gravity used to spot it. In fact, when curvature is present in the membrane (say $K = - 1/a^2$), we have the second scale, $a$, but that is not really necessary as it is $\ell_L$ that identifies the scale at which the continuum field theory description breaks down, opening the doors to the emergence of granular/discretness effects. Such effects are there even when curvature effects are absent ($a \to \infty$). Indeed, in \cite{Iorio:2017vtw,Iorio:2019cvd} it was shown how naturally a GUP emerges in $d=3$ Dirac materials, already in the flat case, when the effects of a nonzero $\ell_L$ are taken into account. On this crucial point, are illuminating the results of Ref.\cite{Jizba:2009qf}, where the fundamental commutator $[\hat{x},\hat{p}]$ has been computed (for the first time) on a generic euclidean lattice.

\section{Impact of the GUP on the BTZ black hole temperature and entropy} \label{BTZHawking}

Armed with the previous results, we want now to focus on how the GUP affects the Hawking temperature and Bekenstein entropy of a \textit{macroscopic} BTZ black hole
\footnote{Two warnings are important here. First, we shall use the GUP in (\ref{3gup_pl}), hence our choice for the gravitational radius in $d=3$ is the event horizon of a \textit{microscopic} BTZ black hole, as given in (\ref{chosenRg}). Second, as in any dimension, also in this case we should not get confused about the logic of having, so to speak, ``two kinds of black-holes'', one microscopic, one macroscopic. In fact, as explained in some details in Section \ref{UPandGrav}, the microscopic black hole is only there associated to the process of localization of a particle with an uncertainty of $\Delta x$, through a photon beam of energy $E$. Such energy can create a gravitational instability (``collapse'') characterized by the event horizon of a micro-black hole with equivalent mass $M = E / c^2$.} in $d=3$.

In fact, we can rewrite formula (\ref{3gup_pl}), by safely assuming the dispersion relation $\Delta E = c\Delta p$, as
\begin{equation}
\Delta x\, \Delta E \geq \frac{\hbar c}{2} \left[1 +  4\beta\left(\frac{\Delta E}{m_{p}c^{2}}\right)^{3/2}\right]\ .
\label{Egup}
\end{equation}
Following \cite{FS9506,AdSTemp,Scardigli:2018jlm,ScardGravTest}, we now first recall how to compute the standard Hawking temperature from the standard HUP, for a $d=4$ Schwarzschild black hole. Then we shall apply the very same technique to obtain the standard Hawking temperature of the $d=3$ BTZ black hole, through the standard HUP (that is the $\beta \to 0$ limit of Eq.(\ref{Egup})). Finally, using the full GUP of (\ref{Egup}), we shall obtain the corrections to the BTZ Hawking temperature for a nonzero $\beta$.

Suppose we are in a $d=4$ spacetime region of weak field (for example, far outside a Schwarzschild black hole), where an effective potential can be defined. Then for any metric of the form $ds^2 = F(r)c^2dt^2 - g_{ik}dx^idx^k$ (where $r^2 = x_1^2 + x_2^2 + x_3^2$) the effective potential reads (see e.g. Refs.\cite{Weinberg1972}, \cite{Norway})
\be
V(r) = \frac{1}{2}c^2(F(r)-1) \; .
\ee
Note that this expression holds as well in a weak field of a $d=3$ spacetime region.
The potential energy of a particle of rest mass $m$ in that region is $U=mV=(F(r)-1) mc^2 /2$. If the particle falls radially in the gravity field for a small radial displacement $\Delta r$, the variation of its potential energy is
\be
\Delta U = \frac{1}{2}mc^2 F'(r)\Delta r \; .
\ee
Suppose that this energy is sufficient to create some particles of mass $m$ from the quantum vacuum, then we can write $\frac{1}{2}mc^2 F'(r)\Delta r = N mc^2$, where $N$ is a form factor related to the particle creation process.
The $\Delta r$ needed for such a process is
\be
\Delta r = \frac{2N}{F'(r)} \; .
\label{Dr}
\ee
The particles so created are confined in a space slice $\Delta r$, so each of them has an uncertainty in energy given by (HUP)
\be
\Delta E \simeq \frac{\hbar c}{2 \Delta r} = \frac{\hbar c}{4 N} F'(r) \; .
\label{DE}
\ee
Interpreting this uncertainty as due to a thermal agitation energy, and using the Maxwell-Boltzmann statistics, we can write the equipartition theorem as
\be
\frac{3}{2}k_B T_{HUP} = \Delta E \simeq \frac{\hbar c}{4 N} F'(r) \; ,
\label{equipartition}
\ee
where $T_{HUP}$ is the temperature of this gas of particles. Therefore
\be
T_{HUP} \simeq \frac{\hbar c}{6 N k_B} F'(r)\,.
\label{TH}
\ee
For a $d=4$ Schwarzschild spacetime $F(r)=1-R_g/r$, with $R_g=2G_NM/c^2$,
and (\ref{TH}) computed at the horizon $r=R_g$ yields
\be
T_{HUP} \simeq \frac{\hbar c}{6Nk_B}\frac{1}{R_g}  = \frac{1}{12 N}  \frac{\hbar c^3}{k_B G_N M} \equiv \frac{2 \pi}{3 N}  \, T_H \; ,
\ee
where the last expression matches the well known Hawking temperature of a $d=4$ Schwarzschild black hole, $T_H \equiv \hbar c^3 / (8 \pi k_B G_N M)$, if we adjust the free parameter $N$ as
$N=2\pi/3$.

We can now repeat a similar argument for the non-rotating BTZ black hole in $d=3$.
From Eq.(\ref{btzcarlip}), with $J=0$, we have $F(r)=(r^2 - R_g^2) / \ell^2$,
with $R_g=\ell\sqrt{8GM}/c$. Using again the standard HUP for the radial coordinate, $\Delta E \simeq \hbar c/(2\Delta r)$, and equation \eqref{Dr}, the equipartition of energy now reads
\be
k_B T_{HUP} = \Delta E \simeq \frac{\hbar c}{4 N} F'(r) \; ,
\label{35}
\ee
where we accounted for the fact that in $d=3$ the spatial degrees of freedom are two, rather then the three of $d=4$. Evaluating (\ref{35}) at the horizon, $r=r_+=R_g = (\ell/c) \sqrt{8G M}$, we get
\be \label{Thup}
T_{HUP} \simeq  \frac{\hbar c R_g}{2N \ell^2 k_B} = \frac{\hbar \sqrt{8G M}}{2 N \ell k_B} \equiv \frac{\pi}{N} \, T_H \,,
\ee
where again, by choosing now $N=\pi$, the last expression matches
\be
T_H \ \equiv \frac{\hbar c R_g}{2 \pi \ell^2 k_B} = \ \frac{\hbar \sqrt{8G M}}{2\pi \ell k_B} \  \; ,
\label{Tbtz}
\ee
which is the well known Hawking temperature of a BTZ black hole (see, e.g., \cite{Carlip2005}).

From the latter expression for the temperature $T_H$, and from the total energy on the hole, $E = Mc^2 = (c^4/\ell^2 8 G) R_g^2$, it is easy to recover the Bekenstein-Hawking entropy of a BTZ black hole, by integrating the thermodynamic definition $dS_{BH}=dE/T_H$. In fact we get
\be\label{Sbh}
S_{BH} = \frac{k_{B} c^{3}}{\hbar G} \; \frac{1}{4} \, ( 2\pi R_g )\;,
\ee
which, in proper units, is the expected one quarter of the $d=3$ black-hole horizon area,
$S_{BH} = {\cal A}/4 $.

We are now ready to compute the corrections to (\ref{Tbtz}) due to the GUP. As first step, consider the inequality (\ref{Egup}) at the saturation,
\be
\Delta E \simeq \frac{\hbar c}{2 \Delta r} \left[1 + 4\beta \left(\frac{\Delta E}{m_{p} c^{2}}\right)^{3/2}\right] \,,
\label{38}
\ee
where in (\ref{Egup}) we choose $x$ to be the radial coordinate $r$, and solve it for
$\Delta E$ as a function of $\Delta r$.
Since the second term in the square brackets is small compared to one, we just need a solution of (\ref{38}) only to first order in $\beta$. In other words, in the second term in the squared brackets we shall use $\Delta E \simeq  \hbar c / (2 \Delta r)$, to obtain
\be
\Delta E \simeq \frac{\hbar c}{2 \Delta r}
\left[1 + 4\beta \left(\frac{\hbar}{2 m_{p} c \Delta r}\right)^{3/2}\right] \;.
\ee
Inserting now $\Delta r$ from equation (\ref{Dr})
and proceeding as before, (cf. Eq. (\ref{35})), we arrive at
\be
 k_B T_{GUP} = \Delta E \simeq \frac{\hbar c}{4N} F'(r)
\left[1 + 4\beta \left(\frac{\hbar F'(r)}{4N m_{p} c}\right)^{3/2}\right]  \;.
\ee
Evaluating this expression at the horizon, $F'(R_g) = 2 R_g / \ell^2$, and following the same logic as above (cf. Eq. (\ref{Thup})), we can write
\be
 T_{GUP} \simeq \frac{\hbar c R_g}{2 N \ell^2 k_B}
 \left[1 + 4\beta \left(\frac{\hbar R_g}{2 N \ell^2 m_{p} c}\right)^{3/2} \right] \,.
\label{TGUP}
\ee
We can fix the free parameter $N$ by demanding the matching of Eq.\eqref{TGUP} with the exact BTZ Hawking temperature \eqref{Tbtz} in the semiclassical limit $\beta \to 0$. So we get $N=\pi$ and finally
\be
T^{(\beta)}_H \equiv T_H \left[1 + 4\beta \left(\frac{\hbar R_g}{2 \pi \ell^2 m_{p} c}\right)^{3/2}\right] \;,
\ee
with the usual $T_H$ given in (\ref{Tbtz}).

Finally, according to the same arguments that lead to entropy $S_{BH}$ in (\ref{Sbh}), it is quite easy to write the GUP-corrected version of the Bekenstein-Hawking entropy for the BTZ black hole. In fact, using $dS^{(\beta)}_{BH}=dE/T^{(\beta)}_H$, to first order in $\beta$ we obtain
\be
S^{(\beta)}_{BH} = S_{BH} \, \left[1 - \frac{8}{5} \beta \left(\frac{\hbar R_g}{2 \pi \ell^2 m_{p} c}\right)^{3/2} \right] \,,
\ee
which is smaller than $S_{BH}$. A comment is in order here. Notice that we find a power law correction to the Bekenstein-Hawking entropy $S_{BH}$, instead of a more common $\log(S_{BH})$ term. But actually, according to what we see in literature
(see e.g.~\cite{Cavaglia:2004jw,Nouicer:2007jg,Glimpses}) about semiclassical corrections to
$S_{BH}$, it is clear that leading $\log(S_{BH})$ term corrections due to GUP appear specifically in $d=4$ dimensions. As soon as we consider GUP corrections in $d \geq 5$ dimensions, the leading terms always follow a power law. So, a leading $\log$-term seems to be a specific feature of four dimensions. Therefore, it doesn't sound surprising that in $d=3$ we find a correction with a power-law leading term.

%
\section{Perspectives and Conclusions}

The various generalizations of the HUP, over the years, have all converged on the idea that the effects of gravity instabilities, caused by a highly energetic process of quantum measurement, must be taken into account. Such gravity-induced GUPs have been extended to dimensions higher than four, but not to lower dimensions, $d=3$ and $d=2$.

Due to the central role played by lower dimensional physics in various contemporary theoretical investigations (from holography in quantum gravity, to dimensional reduction in early cosmology, from the bulk-gravity/boundary-gauge correspondences, to lower dimensional analogs of black hole physics), we intended to fill the gap in this paper. The focus here is on the more straightforward case of $d=3$, although we do point to the main issues of the $d=2$ case, leaving to a later work to address the open questions.

The study revealed to be much more than a mere dimensional analysis of the existing higher dimensional formulae. This is due to the well-known radically different behavior of key geometric tensors, in lower as compared to higher dimensions. In particular, we had to face here the decoupling between Newtonian and Einstein gravity in lower dimensions, that do not allow for a consistent definition of the gravitational radius $R_g$ from Newtonian gravity, as opposed to what happens in $d \ge 4$.

We found, though, that the event horizon of the $M \neq 0$ BTZ micro black hole, that is solution of the $d=3$ Einstein equations with a negative cosmological constant, can be safely taken as the most consistent $R_g$. This gave us the tools to build-up a suitable formula for the $d=3$ GUP we were chasing. We then used the latter formula to estimate the impact of the GUP on the Hawking temperature and Bekenstein entropy of the BTZ black hole.

Taking advantage of the peculiarities of the BTZ black hole, we also pointed here to the extremal $M=0$ case. This approach furnishes an alternative way to the emergence of a maximal resolution/minimal length, in the form of\footnote{During the peer review process, we became aware that a similar proposal already emerged in the discussion of the entanglement entropy of the BTZ black hole, see \cite{BTZzeroM}. There, the AdS length, $\ell$, is promoted to the typical length below which spatial quantum correlation are traced out.} $\ell = 1 / \sqrt{-\Lambda}$. Notice that no such a thing is possible for a standard $d=4$ Schwarzschild black hole, simply because there is no cosmological constant from which one could obtain a second length scale, the first being the spacetime curvature.

This $\ell$ is a possible alternative to the event horizon, to play the role of $R_g$. Here we did not pursue this road till the formulation of a general GUP, but presented instead a specific condensed matter analog realization of this scenario on Dirac material. There $\ell$ emerges as the lattice constant, $\ell_L$, and specific forms of the GUP based on such $\ell$ have been obtained elsewhere, and here just recalled. Notice that the logic for which $\ell_L$ could play the role of a minimal length is somehow complementary to the one involving the formation of micro black holes in the localization process: At those length scales, the standard gravity description, including the smooth manifolds, breaks, in favor of a granular fully quantum description. The famous \textit{space-time foam} envisioned by John Wheeler in the Fifties.

To close, let us point to some of the possible future investigations. As said earlier, surely one direction is to move to $d=2$, and consider the vast family of models with black holes solutions, that should give different $R_g$'s for different models. This is a delicate work that needs be done really case by case, because each case is a different theory of gravity, and we have extensively commented here on how this could affect a proper definition of a $R_g$. Another direction is to consider different $d=3$ gravity theories than the one that is home of the BTZ black hole. One possibility is topologically massive gravity, and its various limiting cases, with or without a cosmological constant. Yet another direction is to include non commutativity of spatial coordinates, $[x_\mu, x_\nu] = i \theta_{\mu \nu}$, in the scenario. Finally, on a more phenomenology tune, all of the above mentioned directions could find experimental realizations in analog gravity models, where dimensionality is often lower than four, one key example being the $d=3$ Dirac materials.

\section*{Acknowledgments}

A.~I. is partially supported by grant UNCE/SCI/013.
P.~P. is supported by the project High Field Initiative (CZ.02.1.01/0.0/0.0/15\_003/0000449) from the European Regional Development Fund.

%
%
%
%
%
%
%

\bibliographystyle{apsrev4-1}
\bibliography{libraryGUP}

\end{document}